\documentclass[a4paper, 10pt, conference]{ieeeconf}     
\IEEEoverridecommandlockouts
\usepackage{amsmath,amsfonts}
\usepackage{algorithmic}
\usepackage{algorithm}
\usepackage{array}
\usepackage[caption=false,font=footnotesize,labelfont=rm,textfont=rm]{subfig}
\usepackage{textcomp}
\usepackage{stfloats}
\usepackage{url}
\usepackage{verbatim}
\usepackage{graphicx}
\usepackage{cite}
\usepackage{multirow}
\UseRawInputEncoding
\begin{document}
\title{\LARGE \bf
Pedestrian Recognition with Radar Data-Enhanced Deep Learning Approach Based on Micro-Doppler Signatures
}

\author{ Haoming Li$^{1}$, Yu Xiang$^{2}$,~\IEEEmembership{Member,~IEEE,} Haodong Xu$^{3}$, and Wenyong Wang$^{4}$,~\IEEEmembership{Member,~IEEE}
\thanks{This work is supported by the National Natural Science Foundation of China (NSFC) Grant no. 62250067.}
\thanks{$^{1}$H. Li, is with School of Computer Science and Engineering, University of Electronic Science and Technology of China (UESTC), Chengdu 611731, China, e-mail: hmli@std.uestc.edu.cn}
\thanks{$^{2}$Y. Xiang, associate professor, is with School of Computer Science and Engineering, UESTC, Chengdu 611731, China, e-mail: jcxiang@uestc.edu.cn}
\thanks{$^{3}$H. Xu, is with School of Computer Science and Engineering, UESTC, Chengdu 611731, China, e-mail: haodongxu@std.uestc.edu.cn}
\thanks{$^{4}$W. Wang, professor, is with School of Computer Science and Engineering, UESTC, Chengdu 611731, China, e-mail: wangwy@uestc.edu.cn}
}
\maketitle

\begin{abstract}
As a hot topic in recent years, the ability of pedestrians identification based on radar micro-Doppler signatures is limited by the lack of adequate training data. In this paper, we propose a data-enhanced multi-characteristic learning (DEMCL) model with data enhancement (DE) module and multi-characteristic learning (MCL) module to learn more complementary pedestrian micro-Doppler (m-D) signatures. In DE module, a range-Doppler generative adversarial network (RDGAN) is proposed to enhance free walking datasets, and MCL module with multi-scale convolution neural network (MCNN) and radial basis function neural network (RBFNN) is trained to learn m-D signatures extracted from enhanced datasets. Experimental results show that our model is 3.33\% to 10.24\% more accurate than other studies and has a short run time of 0.9324 seconds on a 25-minute walking dataset.
\end{abstract}

\begin{keywords}
Pedestrian identification, micro-Doppler effect, data enhancement, multi-characteristic learning (MCL).
\end{keywords}
  
\section{INTRODUCTION}

Due to the increasing of complexity in today's traffic, the demand on perception and recognition of vulnerable road users such as pedestrians and/or cyclists are growing accordingly \cite{wachtel2022convolutional}. Most recognition methods of vulnerable road users are primarily reliant on optical systems, which will be severely hampered in low-light or adverse weather situations (such as rain or heavy fog) \cite{burkacky2018rethinking},\cite{xiang2022multi}. With the development of low-power mmWave radar, it is feasible to use radars installed on automatic vehicles to recognize vulnerable road targets \cite{zhang2019enhanced}. Researchers have shown that each person's micro-motion (such as their swinging arms, legs, or torsos) will modulate the frequency of the radar echo signal, resulting in sidebands with respect to the person's Doppler frequency shift. This frequency modulation effect is called the micro-Doppler (m-D) effect \cite{chen2006micro}. Each pedestrian produces unique m-D signatures that could be utilized to detect and recognize \cite{xiang2022multi},\cite{nanzer2017review}.

Research on the m-D signatures of human motion was firstly carried out in 1998 \cite{chen2014radar}. Pedestrian recognition has garnered a lot of attention in recent years, with advances both in radar technology and deep learning \cite{tavanti2021short},\cite{hor2022single},\cite{ferguson2018mmwave}. Cao et al. used a K-band Doppler radar and proposed the deep convolutional neural network (DCNN) recognition approach \cite{cao2018radar}. The m-D signatures of 22 people walking on a treadmill were collected using a 25GHz continuous-wave (CW) radar in \cite{abdulatif2019person}. Lang et al. used a plain convolutional neural network (CNN) with a multiscale feature aggregation strategy to identify four walking people \cite{lang2020person}. Signal statistical features, calculated from m-D signatures, were also be used to distinguish humans from other objects or to distinguish different human movements \cite{kim2014human},\cite{seyfiouglu2018deep}. All above researches only considered time-Doppler spectrogram (TDS) or Signal statistical features as m-D signatures, thus the micro-motion information might not be fully learned. In addition, within these researches, pedestrians' movements were constrained, that means they only approached or stayed away from the radar, or walked with pre-determined route.

Researches on pedestrian recognition problems used multiple categories of m-D signatures (i.e. both with TDS and signal statistical features) with uncontrolled movements of pedestrians have just begun. Vandersmissen et al. extracted m-D signatures to identify five people walking freely indoors using FMCW radar \cite{vandersmissen2018indoor}. However, they still only considered TDS data as m-D signatures. A novel deep learning model named multi-characteristic learning (MCL) model for pedestrian identification was proposed in \cite{xiang2022multi}. MCL model jointly learned discrepant pedestrian m-D signatures and fused the knowledge learned from each signature into final decisions and showed its accuracy from 80\%-96\% within free-walking datasets.

Generally, deep learning model is commonly built on a large amount of training data. Lack of adequate training data hampers deep learning model's accuracy \cite{qiao2022person}. Due to high computing power demand and experimental cost, it is not feasible for researchers to acquire adequate training data in radar-based pedestrian recognition area \cite{lang2020person},\cite{weller2022mimo}. In the past few years, generative adversarial networks (GANs), provide a way to learn deep representations that generate new data with the same statistics of the training data, have been adopted in radar signal processing area \cite{creswell2018generative},\cite{rahman2022physics},\cite{alnujaim2021synthesis}.

Inspired by previous works, in this paper, we propose a novel data enhanced deep learning approach to utilize more complementary information for pedestrian recognition. Main contributions of our work are as follows: 1) we propose range-Doppler GAN (RDGAN) to learn spatial-tempo relationship among time-Doppler spectrograms and generate new TDS data, thus the training datasets are expanded. 2) a novel MCL model with multi-scale convolution neural network (MCNN) and radial basis function neural network (RBFNN) as function networks is proposed to learn micro-motion information within various pedestrians. 3) finally, a novel deep learning approach combine with RDGAN and MCL module, named data-enhanced multi-characteristic learning (DEMCL) model, is proposed and validated with free walking datasets.

The remainder of this paper is structured in the following manner. In section II, we propose and describe DEMCL model to learn m-D signatures. Section III compares and analyzes the results of experiments. The paper is concluded in section IV.
\section{System Description}
\begin{figure}[!ht]
	\centering
	\includegraphics[width=3.4in]{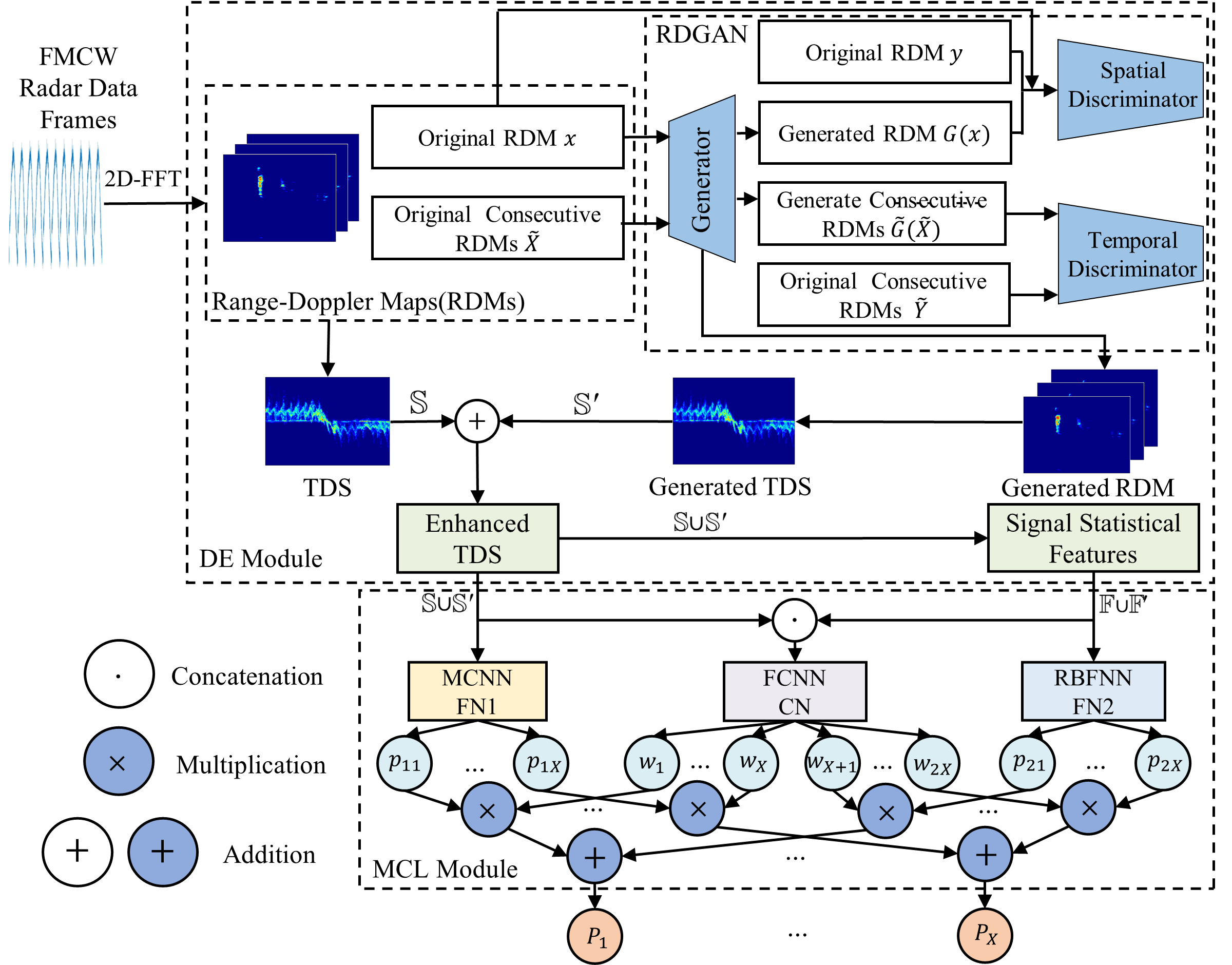}
	\caption{Block Diagram of DEMCL}
	\label{fig_1}
\end{figure}

In this section, we propose a novel data enhanced deep learning approach named DEMCL model. DEMCL model is composed of Data Enhancement (DE) module and MCL module (Fig.\ref{fig_1}). In DE module, a spatial-tempo relative GAN named RDGAN is proposed to generate new TDS data for data enhancement. Two categories of m-D characteristics (i.e. TDS and signal statistical features) are acquired by processing generated TDS and original TDS in DE module and input into MCL module. MCL module is composed of two function networks (FN1 and FN2) and a context network(CN). FN1 uses TDS to recognize pedestrians, and its structure is MCNN; FN2 uses signal statistical features to recognize pedestrians, and its structure is RBFNN; m-D signatures are fused into CN to learn weights that two m-D signatures influence on final decisions. 
\vspace{-0.3cm}
\vspace{-0.8cm} 
\subsection{Data Enhancement Module}
From original time-varying frequency radar echo signal, range-Doppler maps (RDMs) and TDS are calculated.  
\subsubsection{Time-Doppler spectrogram}
We assume that signal $s(k, l)$ of size $K \times L$ forms a frame, where $K$ denotes the number of sampling points in a chirp and $L$ denotes the number of chirps contained in a FMCW radar data frame. Through 2D fast Fourier transform (2D-FFT), RDM $S(u,v)$ is acquired as follows:
\begin{equation}
	\label{equa_2}
	S(u,\!v)\!=\!\sum_{l=0}^{L-1}\!\sum_{k=0}^{K-1}\!s(k,l)e^{-j 2\pi\left(\frac{uk}{K}+\frac{vl}{L}\right)},(u,v\!=\!1\!,\cdots\!,K\mid L)
\end{equation}
$e_v$ donates the value of each Doppler cell element in RDM, and it is represented in decibels in frequency domain. After RDMs acquired, $e_v$ are added together along range axis to form a vector $\mathbf{e}$ (i.e. L Doppler cells) for each RDM, as described in (\ref{equa_3}). Finally, $\mathbf{E}$ is obtained by combining $n$ time-continuous $\mathbf{e}$, which forms a TDS, given by (\ref{equa_3}):
\begin{equation}
	\label{equa_3}
 \setlength{\arraycolsep}{0pt}
	\left\{\begin{array}{c}
		e_v=\sum_{u=1}^{K} 20 \log _{10}|S(u, v)|,(u=1, \cdots, K) \\ \\
   \setlength{\arraycolsep}{0pt}
		\mathbf{e}_i=[{e_1},\cdots,{e_v},\cdots,{e_L}]_i^{\mathrm{T}},(v,i=1, \cdots, L|n)\\ \\
		\mathbf{E}=[\mathbf{e_1}, \mathbf{e_2}, \cdots, \mathbf{e_n}]
	\end{array}\right.
\end{equation}
After that, noise is filtered out in TDS. Details are shown in\cite{xiang2022multi}.
\subsubsection{Range Doppler Generative Adversarial Network}
\begin{figure}[!ht]
 \setlength{\abovecaptionskip}{0.cm}
 \vspace{-0.2cm}
	\centering
	\includegraphics[width=3.4in]{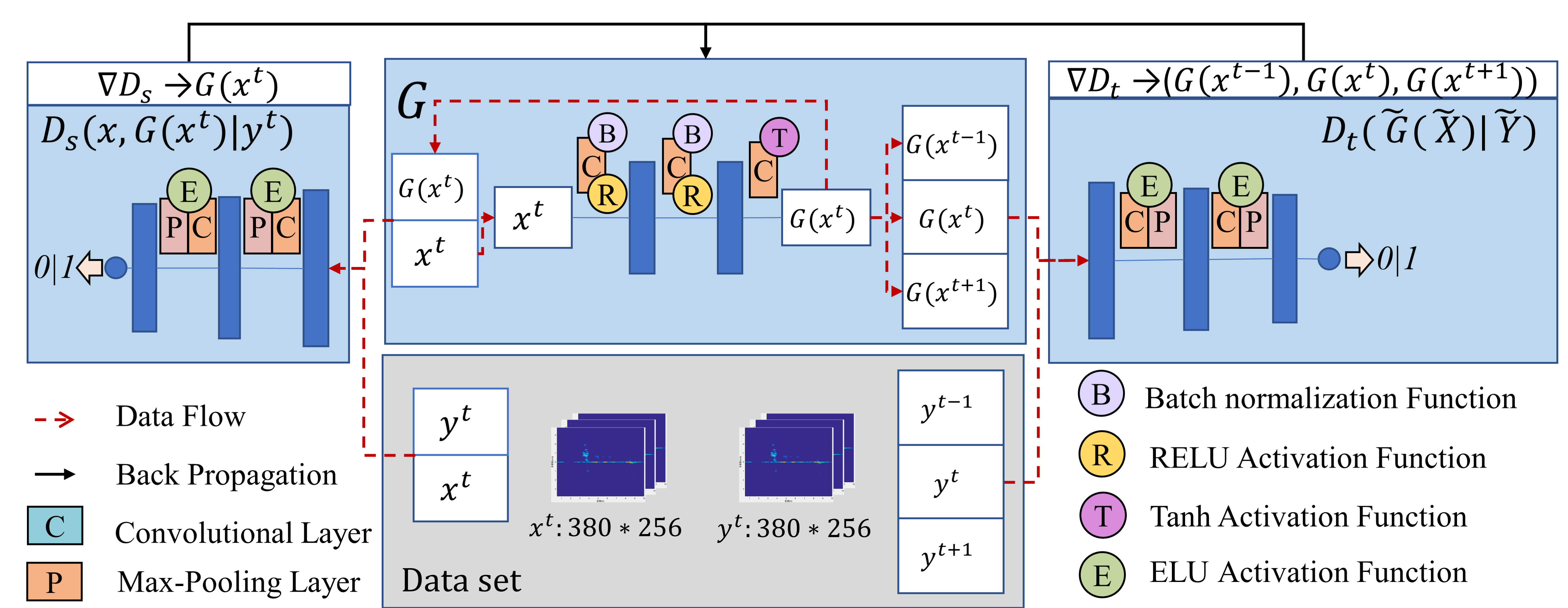}
	\caption{The structure of RDGAN}
	\label{fig_2}
 \vspace{-0.2cm}
\end{figure}

In order to learn spatial-tempo relationship among TDSs and generate new TDS data, inspired by tempoGAN \cite{xie2018tempogan}, we propose RDGAN which is composed of a generator $G$, a spatial discriminator $D_s$ and a temporal discriminator $D_t$ (Fig.\ref{fig_2}). Original RDMs are acquired in (\ref{equa_2}). Then the skew-normal distributed noise is filtered out and signal strength of the zero Doppler channel is reduced in RDMs. Time-continuous $n$ frames of RDMs are used as input of RDGAN. In Fig.\ref{fig_2}, $x^t$ represents frame $t$ of original RDMs, $y^t$ represents frame $t+1$ of original RDMs and $G(x^t)$ represents frame that $G$ generates with input $x^t$($t=2,3,\cdots,n-2$). $\tilde{X}$ represents $\left\{x^{t-1},x^t,x^{t+1}\right\}$; $\tilde{Y}$ represents $\left\{y^{t-1},y^t,y^{t+1}\right\}$; $\tilde{G}(\tilde{X})$ represents $\left\{G(x^{t-1}),G(x^t),G(x^{t+1})\right\}$.

During one training iteration of RDGAN, firstly, $D_s$, a conditional discriminator realized as a CNN, is trained to reject $G(x^t)$ (i.e. $D_s(x^t,G(x^t))\!\rightarrow\!0$) and accept $y^t$ (i.e. $D_s(x^t,y^t))\!\rightarrow\!1$). The loss for $D_s$ is:
\begin{equation}
	\small
 \label{equa_4}
        \begin{aligned}
            L_{D_{s}}\!(\!D_s,\!G)\!=\!(\!-\mathbb{E}_m\![logD_s(x^t\!,\!y^t\!)]\!-\!\mathbb{E}_m[log(1\!-\!D_{s}(x^t\!,\!G(x^t)\!)\!)])/2     
        \end{aligned}
\end{equation}
where $m$ denotes the number of samples. $D_s$ is updated with $\nabla_{D_s}[L_{D_s}(D_s,G)]$ for $k_{D_s}$ times.

Then, $D_t$, an unconditional discriminator realized as a CNN, is trained to reject temporal changes in $\tilde{G}(\tilde{X})$ (i.e. $D_t(\tilde{G}(\tilde{X}))\!\rightarrow\!0$) and accept temporal changes in $\tilde{Y}$ (i.e. $D_t(\tilde{Y})\!\rightarrow\!1$). The loss for $D_t$ is:
\begin{equation}
	\label{equa_6}
        \begin{aligned}
            L_{\!D_t}\!(\!D_t,G)\!=\!(\!-\mathbb{E}_m[logD_t(\tilde{Y}\!)]\!-\!\mathbb{E}_m[log(1\!-\!D_t(\tilde{G}(\tilde{X}))\!)])/2 
        \end{aligned}
\end{equation}
$D_t$ is updated with $\nabla_{D_t}[L_{D_t}(D_t,G)]$ for $k_{D_t}$ times. 

Finally, $G$, a generator realized as a fully convolutional network (FCN), is trained to generate $G(x^t)$ and $\tilde{G}(\tilde{X})$ which could be accepted by $D_s$ and $D_t$ (i.e. $D_s(x^t,G(x^t))\!\rightarrow\!1$, $D_t(\tilde{G}(\tilde{X}))\!\rightarrow\!1$). The loss for $G$ is:
\begin{equation}
\begin{small}
	\label{equa_8}
        \begin{aligned}
            L_G(\!D_s,\!D_t,\!G)\!=\!(\!-\!\mathbb{E}_m[logD_s(x^t\!,G(x^t))]\!-\!\mathbb{E}_m[logD_t(\tilde{G}(\tilde{X}))])\!/2
        \end{aligned}
\end{small}
\end{equation}

$G$ is updated with $\nabla_G[L_G(D_s,D_t,G)]$ for $k_G$ times.

The process above is repeated until RDGAN converges.

Complete model training strategy is described in Algorithm 1. $epochs$ denotes the number of training iterations.

After training, new RDMs is generated by trained $G$. Then generated TDS data $\mathbf{E}'$ is obtained as the same as $\mathbf{E}$ in (\ref{equa_3}).

\floatname{algorithm}{Algorithm}
\renewcommand{\algorithmicrequire}{\textbf{Initialization:}}
\renewcommand{\algorithmicensure}{\textbf{Input:}}
\begin{algorithm}[H]
	\caption{Training Procedure}
	\begin{algorithmic}[1]
		\REQUIRE $D_s$, $D_t$, $G$         
		\ENSURE $x^t$($G$ and $D_s$ input),  $\tilde{X}$($G$ input), $y^t$($D_s$ input), $\tilde{Y}$($D_t$ Input)
            \renewcommand{\algorithmicrequire}{\textbf{Output}}
            \REQUIRE $G(x^t)$(G output),  $\tilde{G}(\tilde{X})$(G output)
		\FOR{epochs}  
			\FOR{$k_{D_s}$}     
			\STATE \hspace{0.5cm} Compute RDM $G(x^t)$ by $G$
			\STATE \hspace{0.5cm} Update $D_s$ with $\nabla_{D_s}[L_{D_s}(D_s,G)]$
                \ENDFOR
                \FOR{$k_{D_t}$}  
			\STATE \hspace{0.5cm} Compute RDM $G(x^t)$ by $G$
			\STATE \hspace{0.5cm} Update $D_t$ with $\nabla_{D_t}[L_{D_t}(D_t,G)]$
                \ENDFOR
                \FOR{$k_G$} 
			\STATE \hspace{0.5cm} Compute RDM $G(x^t)$ and $\tilde{G}(\tilde{X})$ by $G$
			\STATE \hspace{0.5cm} Update $G$ with $\nabla_G[L_G(D_s,D_t,G)]$ 
			\ENDFOR 
		\ENDFOR  
	\end{algorithmic}
\end{algorithm}

\subsubsection{Signal Statistical Features}
Four features are computed from TDS: torso Doppler frequency $f_1$, bandwidth of the Doppler signal $f_2$, bandwidth of the torso alone $f_3$, and period of the limb motion $f_4$ . These features are calculated from spectrogram in a time window of size $T_z$ that contain Z time-consecutive frames, and details are shown in \cite{xiang2022multi}.

\subsection{Multi-Characteristic Learning (MCL) Module}

MCL module is composed of three parts: two functional networks and a context network.

\begin{figure}[!ht]
 \vspace{-0.3cm}
 \setlength{\abovecaptionskip}{0.cm}
	\centering
	\includegraphics[width=3.4in]{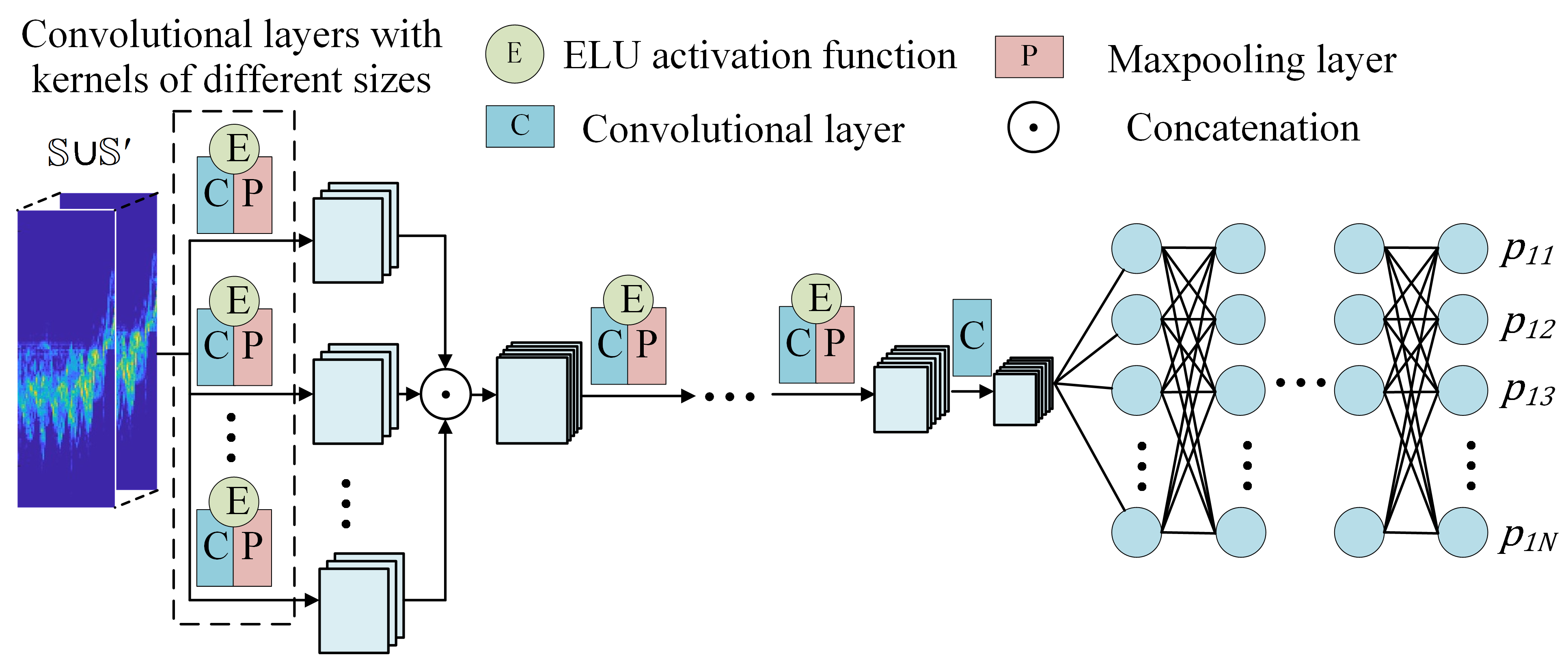}
	\caption{The structure of FN1}
	\label{fig_3}
 \vspace{-0.3cm}
\end{figure}
In \cite{xiang2022multi}, FN1 is designed as a DCNN. The process of characteristic extraction and object classification process are detached in DCNN, thus various characteristics existed in various frequency domains are difficult to be fully extracted. To solve this problem, MCNN with multi-frequency branch and local convolution was proposed in \cite{cui2016multi}. So in our work, FN1 is designed as MCNN with filters of different sizes (Fig.\ref{fig_3}), responsible for processing TDS. Original TDS $\mathbb{S}$ and generated TDS $\mathbb{S}'$ are combined and input into FN1. $\mathbb{S}$ and $\mathbb{S}'$ are obtained as follows:
\begin{equation}
\label{equa_A}
\mathbb{S}=\bigcup_{i=1}^M({\mathbf{E}_{T_z}})_{i}, \mathbb{S}'=\bigcup_{i=1}^N({\mathbf{E}'_{T_z}})_{i}
\end{equation}
$M$ represents the number of samples in $\mathbb{S}$, and $N$ represents the number of samples in $\mathbb{S}'$.

\begin{figure}[!ht]
 \vspace{-0.3cm}
 \setlength{\abovecaptionskip}{0.cm}
	\centering
	\includegraphics[width=2.4in]{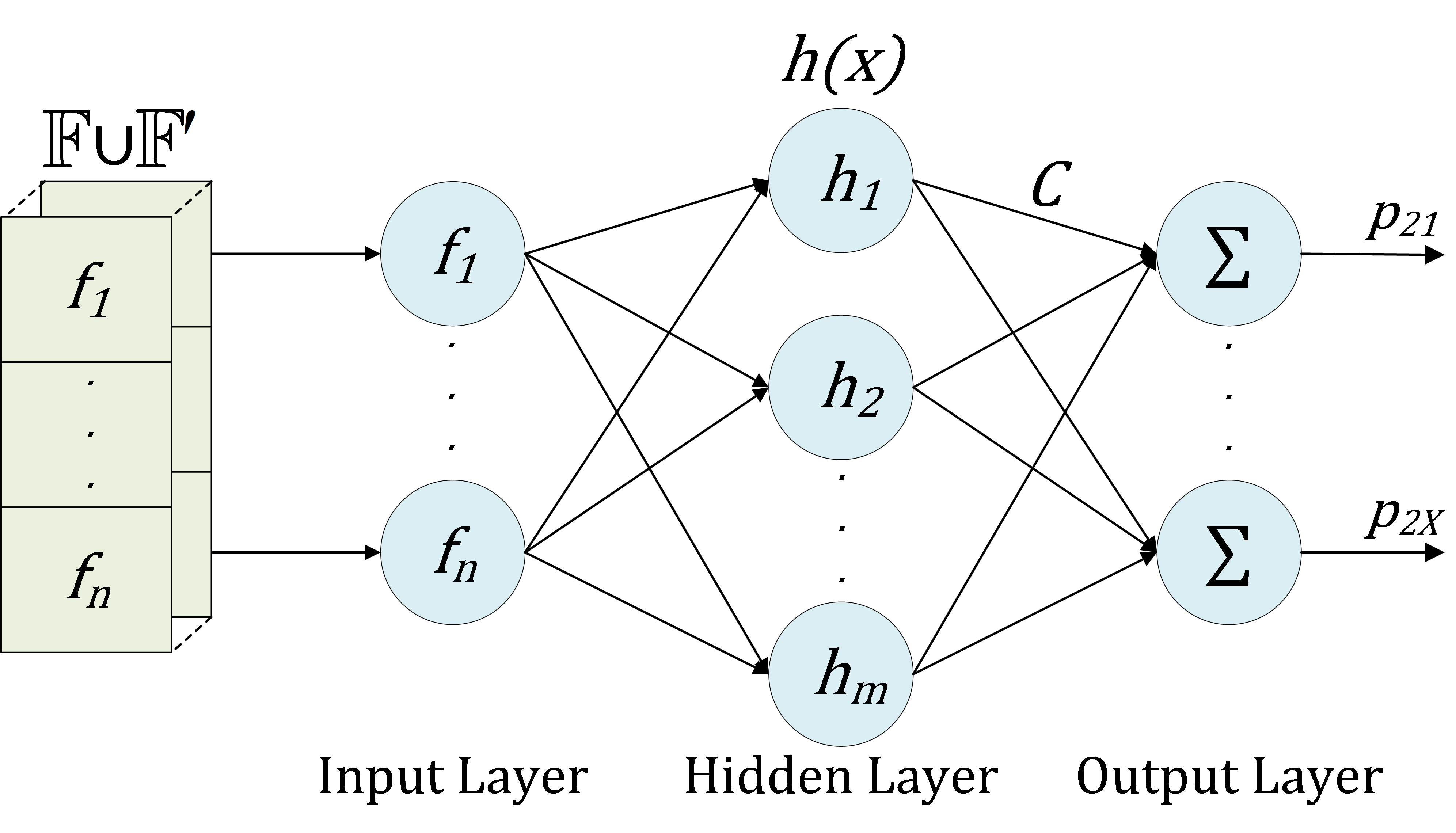}
	\caption{The structure of FN2}
	\label{fig_4}
  \vspace{-0.3cm}
\end{figure}

In \cite{xiang2022multi}, FN2 is designed as a fully connected neural network (FCNN). FCNN does not have the optimal approximation property for any continuous function \cite{girosi1990networks}, based on which, it is proved that a RBFNN is the optimal approximation for given continuous functions, with the help of regularization theory \cite{poggio1990networks}. In addition, Hu et al, expanded above theory and proved that a RBFNN is optimal approximation to a discriminator of GAN\cite{hu2022flow}. Therefore, while giving a continuous function, RBFNN approximates better than FCNN. So FN2 is designed as a RBFNN (Fig.\ref{fig_4}), responsible for processing statistical features. Original signal statistical features $\mathbb{F}$ and generated signal statistical features $\mathbb{F}'$ are combined and input into FN2. $\mathbb{F}$ and $\mathbb{F}'$ are obtained as follows:
\begin{equation}
\label{equa_B}
 \mathbb{F}=\bigcup_{i=1}^{M} [f_1,f_2,\cdots,f_n]_i,  \mathbb{F}'=\bigcup_{i=1}^{N} [f'_1,f'_2,\cdots,f'_n]_i
\end{equation}
where $[f_1,f_2,\cdots,f_n]$ represents $n$ signal statistical features in a $T_z$, and $f'_i$ represents signal statistical features obtained from $\mathbb{S}'$. In Fig.\ref{fig_4}, $C$ is a vector made up of the weight between every hidden neuron and every output neuron. $h(x)$ represents radial basis function (RBF) that is Gaussian kernel function:
\begin{equation}
\setlength\abovedisplayskip{0cm}
	\label{equa_gaussian}
	h(\mathbf{f_k},\mathbf{v_j},{\sigma_j})=\operatorname{exp}(-\frac{{||\mathbf{f_k}-\mathbf{v_j}||}^2}{2{\sigma_j}^2})
\end{equation}
where $||\cdot||$ is a norm on the function space, $\mathbf{f_k}$ denotes all signal statistical features of $k$-th sample in training set (i.e. $\mathbf{f_k}=[f_1, f_2, \cdots, f_n]_k$), and $\mathbf{v_j}$ and $\sigma_j$ denote the center and width of the $j$-th hidden neuron, respectively. The output of FN2 is written as:
\begin{equation}
	\label{equa_FN2}
	p_{2i}=c_{0i}+\sum_{j=1}^{q}c_{ji}h(\mathbf{f_k},\mathbf{v_j},{\sigma_j})
\end{equation}
where $p_{2i}$ represents the probability that input is judged as the $i$-th pedestrian by FN2. $q$ denotes the number of hidden neurons, $c_{0i}$ denotes bias of the $i$-th output neuron, and $c_{ji}$ denotes weight between the $j$-th hidden neuron and the $i$-th output neuron.

CN is a FCNN, responsible for learning $\mathbb{S} \cup \mathbb{S}'$ and $\mathbb{F} \cup \mathbb{F}'$ differentiation and correlations in pedestrian recognition:
\begin{equation}
	\label{equa_10}
	\left\{\begin{array}{c}
		p_{1i}=FN1(\mathbb{S} \cup \mathbb{S}'), i=1,2, \cdots, X \\ \\
		p_{2i}=FN2(\mathbb{F} \cup \mathbb{F}'), i=1,2, \cdots, X\\ \\
		w_{i}\!=\!CN((\mathbb{S}\!\cup\!\mathbb{S}')\!\cup\!(\mathbb{F}\!\cup\!\mathbb{F}')), i=1,2, \cdots, 2X
	\end{array}\right.
\end{equation}
where $X$ represents the number of pedestrians to be identified, $p_{1i}$ represents the probability that input is judged as the $i$-th pedestrian by FN1. $w_i$ represents weight values learned during joint learning procedure of two m-D signatures in CN. $\mathbb{S} \cup \mathbb{S}'$ and $\mathbb{F} \cup \mathbb{F}'$ are combined and input into CN to learn weights of two categories of m-D signatures to obtain more accurate fusion judgment results.

Two FNs and CN are trained together in MCL as follows.

Step1: initialize the network parameters. 

Step2: forward propagation. Cross-entropy loss function is:
\vspace{-0.2cm}
\begin{equation}
\label{equa_11}
\operatorname{CE}(\mathbf{label},\!\mathbf{P})\!=\!\frac{1}{M\!+\!N}\!\sum_{i=1}^{M\!+\!N}\!\left[\!-\!\sum_{j=1}^{X} {label}_{j|i}\!\left(\log\!\left({P}_{j|i}\!\right)\!\right)\!\right]
\end{equation}
\vspace{-0.2cm}
\begin{equation}
\nonumber
label_{i|j}=
\begin{cases} 
1, $The category of j is equal to i$ \\
0, $The category of j is not equal to i$
\end{cases}
\label{eq:label}
\end{equation}

${P}_{j|i}$ is the probability that the $i$-th sample in MCL will be identified as the $j$-th category.

Step3: function networks updates.
Two function networks and context networks' parameters are updated by:
\begin{equation}
 \vspace{-0.1cm}
 \nonumber
	\label{equation_update_function_networks_W_b}
	\mathbf{W}=\mathbf{W}+\gamma \frac{\partial loss}{\partial \mathbf{W}},	\mathbf{b}=\mathbf{b}+\gamma \frac{\partial loss}{\partial \mathbf{b}}
\end{equation}
where, $\gamma$ denotes the learning rate of MCL, $\mathbf{W}$  and  $\mathbf{b}$ denote weight matrix and bias vector of MCL, respectively.
Specially, in functional network 2, two parameters are updated by:
\begin{equation}
 \vspace{-0.2cm}
 \nonumber
	\label{equation_update_function_networks_c_v}
	\mathbf{\sigma}=\mathbf{\sigma}+\gamma \frac{\partial loss}{\partial \mathbf{\sigma}}, \mathbf{V}=\mathbf{V}+\gamma \frac{\partial loss}{\partial \mathbf{V}}
\end{equation}
where $\mathbf{\sigma}$ and $\mathbf{V}$ denote the center matrix and width vector of RBFNN hidden layer.

Step4: repeat step 1, 2, 3 until the MCL converges. 

The output of FN1, FN2 and CN is fused as the final decision output:
\begin{equation}
	\label{equa_1}
	\left\{\begin{array}{c}
		P_i = p_{1i} * w_i + p_{2i} * w_{X+i} \\ \\
		\mathbf{P} =[P_1,\cdots,P_i,\cdots,P_X], (i = 1, \cdots, X)
	\end{array}\right.
\end{equation}
$P_i$ is the probability of being identified as the $i$-th pedestrian. $\mathbf{P}$ is a vector made up of $X$ pedestrian identification probabilities.

\section{Experimental Results And Analysis}
\subsection{Experiment Descriptions}
Open-source data collection IDRad which contains 150 minutes of tagged pedestrian FMCW radar data is adopted in our experiments. The pedestrian radar data were collected while five people with different body types and similar postures walked freely\cite{vandersmissen2018indoor}. 

RDGAN's structure parameters are summarized in Table \ref{table_1} and its structure is shown in Fig.\ref{fig_2}. 6 minutes of continuous walking freely data for each pedestrian is used as original data. In Table \ref{table_1}, CNN $(a, b, c)$ represents a three-layer CNN with number of channels labeled $a$, $b$, $c$. FCNN $(d,e)$ represents a FCNN with number of neurons in each layer from input to output being $d$ and $e$. FCN $(f, g, h, i)$ represents a four-layer FCN with number of channels labeled $f$, $g$, $h$ and $i$.  Input size of RDM is $380\times256$, where 380 denotes the number of range channels (i.e. range from 0.5m to 10m), and the number of Doppler channels is 256. $k_{D_s}$, $k_{D_t}$ and $k_G$ are all set as 1. The learning rate is 0.0005, with a total of 100 epochs.
\vspace{-0.3cm} 
\begin{table}[!ht]
\setlength{\abovecaptionskip}{-0.1cm}
\setlength{\belowcaptionskip}{-0.3cm}
	\caption{RDGAN's comprehensive structure\label{table_1}}
	\centering
        \scriptsize{
		\begin{tabular}{|c|c|c|}
			\hline
			Network & Structure & Kernel\\
			\hline
			\multirow{2}{*}{$D_s$} & CNN (2,8,16) & $3\times3$\\
			\cline{2-2}\cline{3-3}
			& FCNN (5888,1) &\\
			\hline
			\multirow{2}{*}{$D_t$} & CNN (3,8,16) & $3\times3$\\
			\cline{2-2}\cline{3-3}
			& FCNN (5888,1) &\\
			\hline
			$G$ & FCN (1,32,8,1) & $3\times3$\\
			\hline
		\end{tabular}
	}
\end{table}
\vspace{-0.3cm}

MCL module's structure parameters are summarized in Table \ref{table_2} and its structure is shown in Fig.\ref{fig_3}. 25 minutes of walking data and 20 minutes generated data for each pedestrian is used as training set, and use other 5 minutes of walking data as test set. In Table \ref{table_2}, MCNN $(j, k, l)$ represents the number of channels of each convolutional layer labeled $j$, $k$, $l$ respectively in a MCNN layer. RBF$(m,n,o)$ represents a RBFNN with number of neurons in each layer from input to output being $m$, $n$, $o$. Input size of TDS is $45\times205$, where 45 denotes 45 consecutive frames of TDS (i.e. 3-$s$), and the number of doppler channels is 205. In FN2, four signal statistical features from each of the chosen 165 consecutive frames (i.e. $T_z$ = 11-s) are used as input. In CN, its input is a vector that combines inputs of FN1 and FN2. The learning rate is 0.0005, with a total of 500 epochs.
\vspace{-0.3cm} 
\begin{table}[!th]
\setlength{\abovecaptionskip}{-0.1cm}
\setlength{\belowcaptionskip}{-0.3cm}
	\caption{MCL's comprehensive structure\label{table_2}}
	\centering
        \scriptsize{
		\begin{tabular}{|c|c|c|}
			\hline
			Network & Structure & Kernel\\
			\hline
			\multirow{3}{*}{FN1} & MCNN (8,8,8) & $3\times3$, $4\times4$, $5\times5$\\
			\cline{2-2}\cline{3-3}
			& CNN (24,48,96,128,32) & $3\!\times\!3$, $1\!\times\!1$(last layer)\\
   			\cline{2-2}\cline{3-3}
			& FCNN (800,128,5) & \\
			\hline
			FN2 & RBFNN (4,5,5) & \\
			\hline
			CN & FCNN (9229,1000,100,10) & \\
			\hline
		\end{tabular}
	}
\vspace{-0.3cm}
\end{table}

\subsection{Results and Analysis of RDGAN}
\begin{figure}[!ht]
 \centering
	\subfloat[\scriptsize\normalfont Pedestrian 3]{\includegraphics[width=0.24\textwidth]{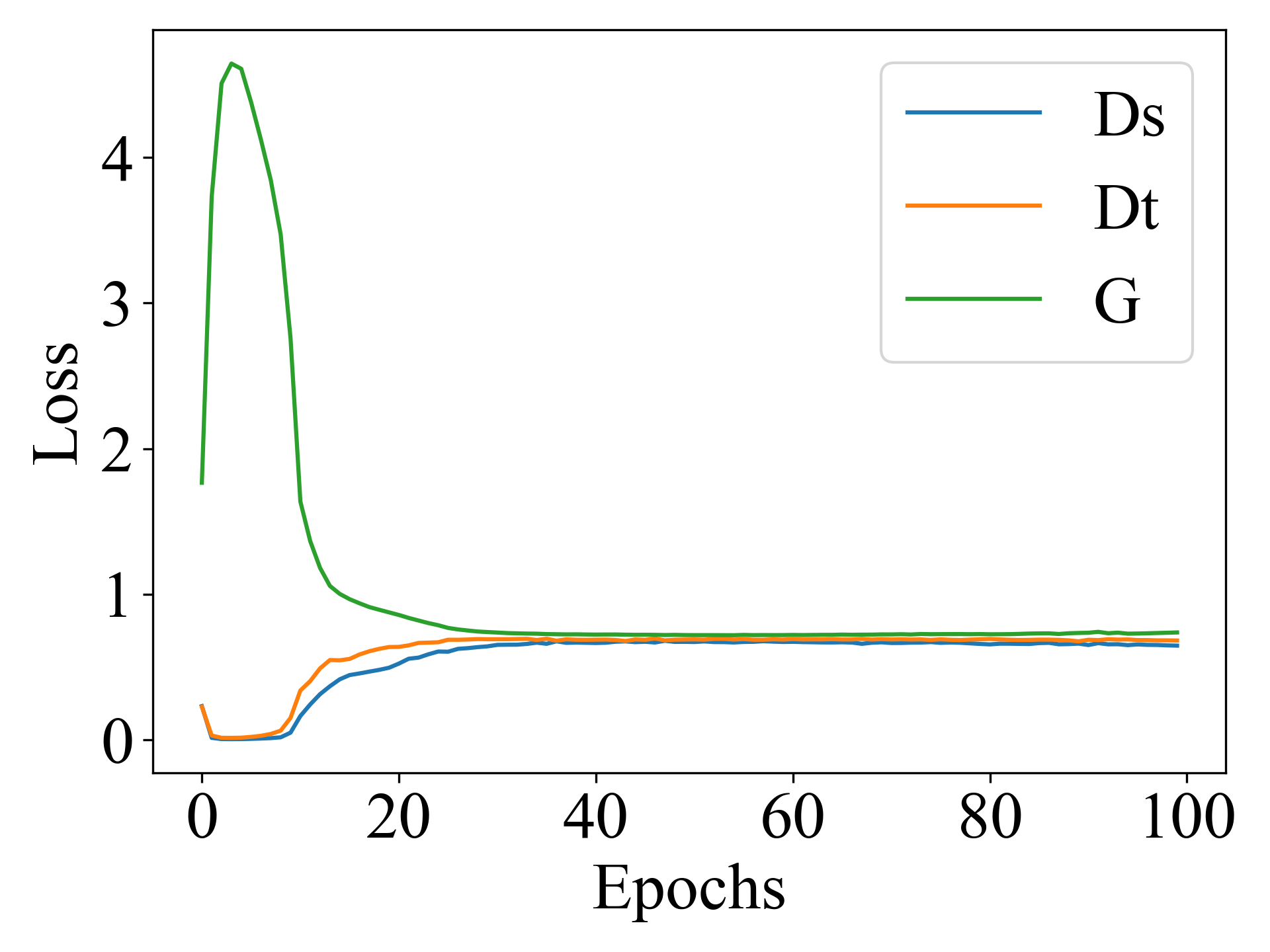}}
	\hfil
	\subfloat[\scriptsize\normalfont Pedestrian 5]{\includegraphics[width=0.24\textwidth]{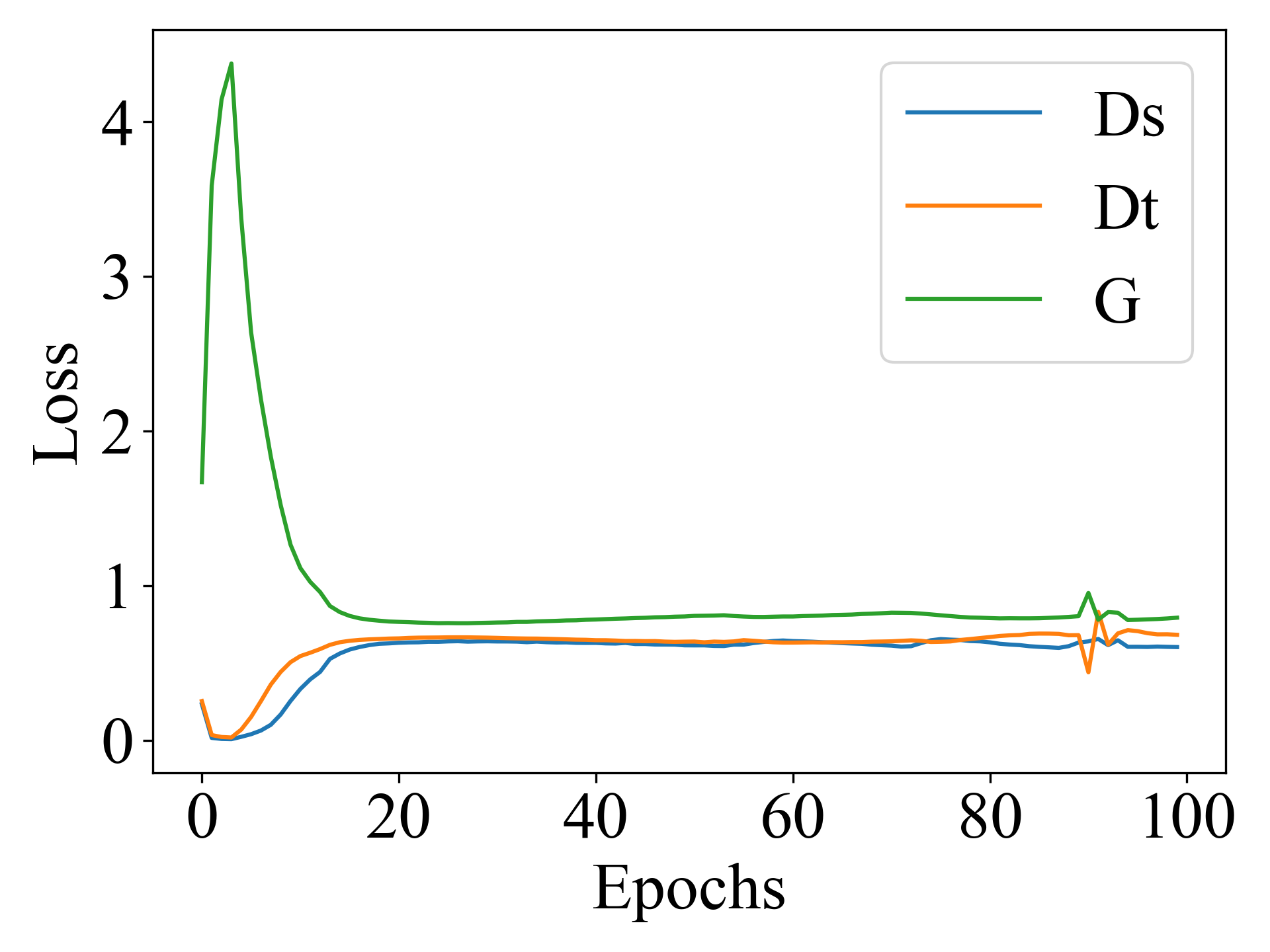}}
	\caption{Loss curves of two pedestrians}
  \vspace{-0.3cm}
	\label{fig_5}
\end{figure}

Loss value curves of RDGAN are shown in Fig.\ref{fig_5}. Two models converge after 20-th epoch. In fact, five models' loss function values of $G$, $D_s$ and $D_t$ all reach a similar value of about 0.73, this means that they all converge.

\begin{figure}[!ht]
 \vspace{-0.4cm}
	\centering
	\subfloat[\scriptsize\normalfont Original RDM]{\includegraphics[width=0.24\textwidth]{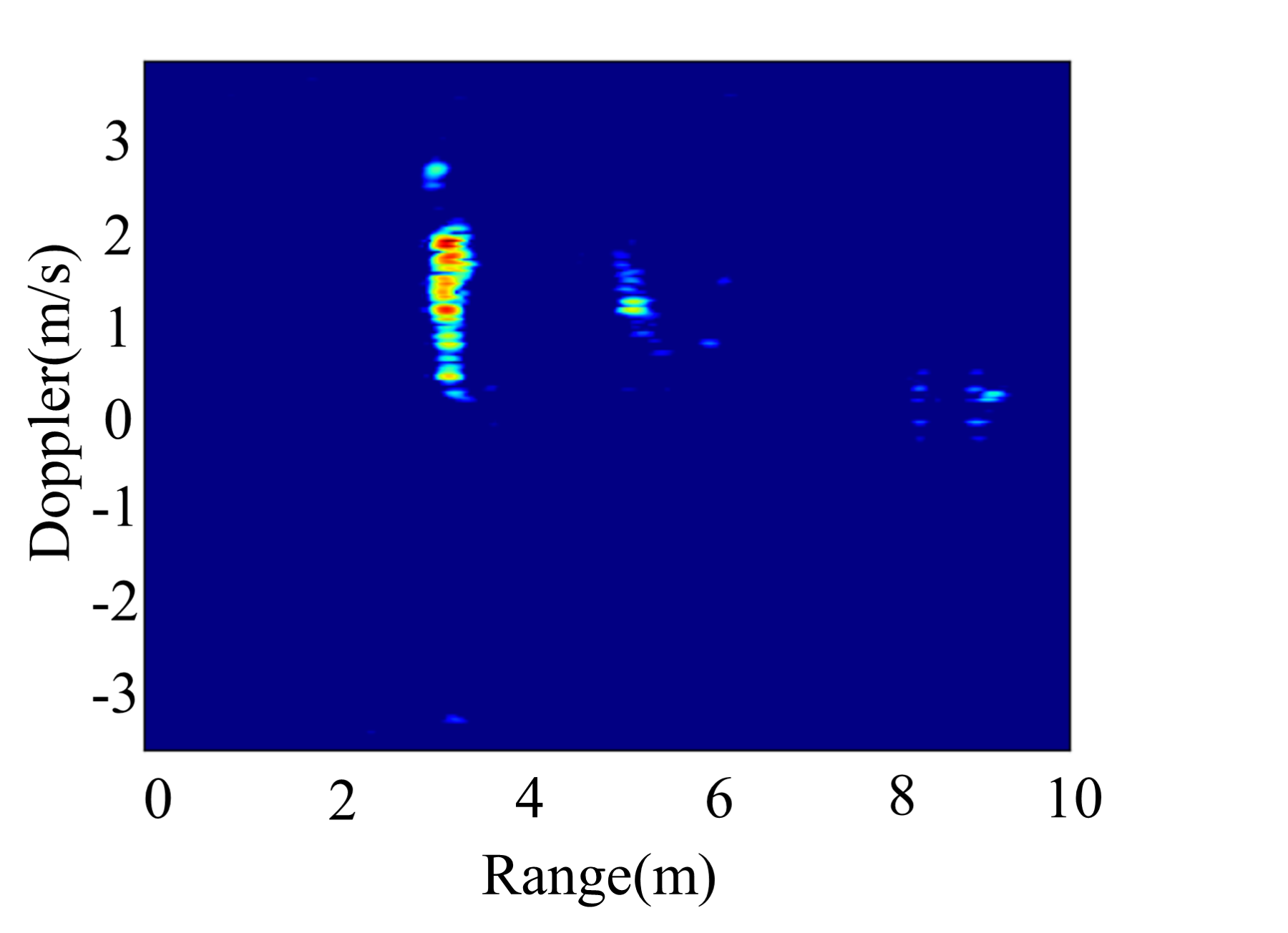}}
	\hfil
	\subfloat[\scriptsize\normalfont Generated RDM]{\includegraphics[width=0.24\textwidth]{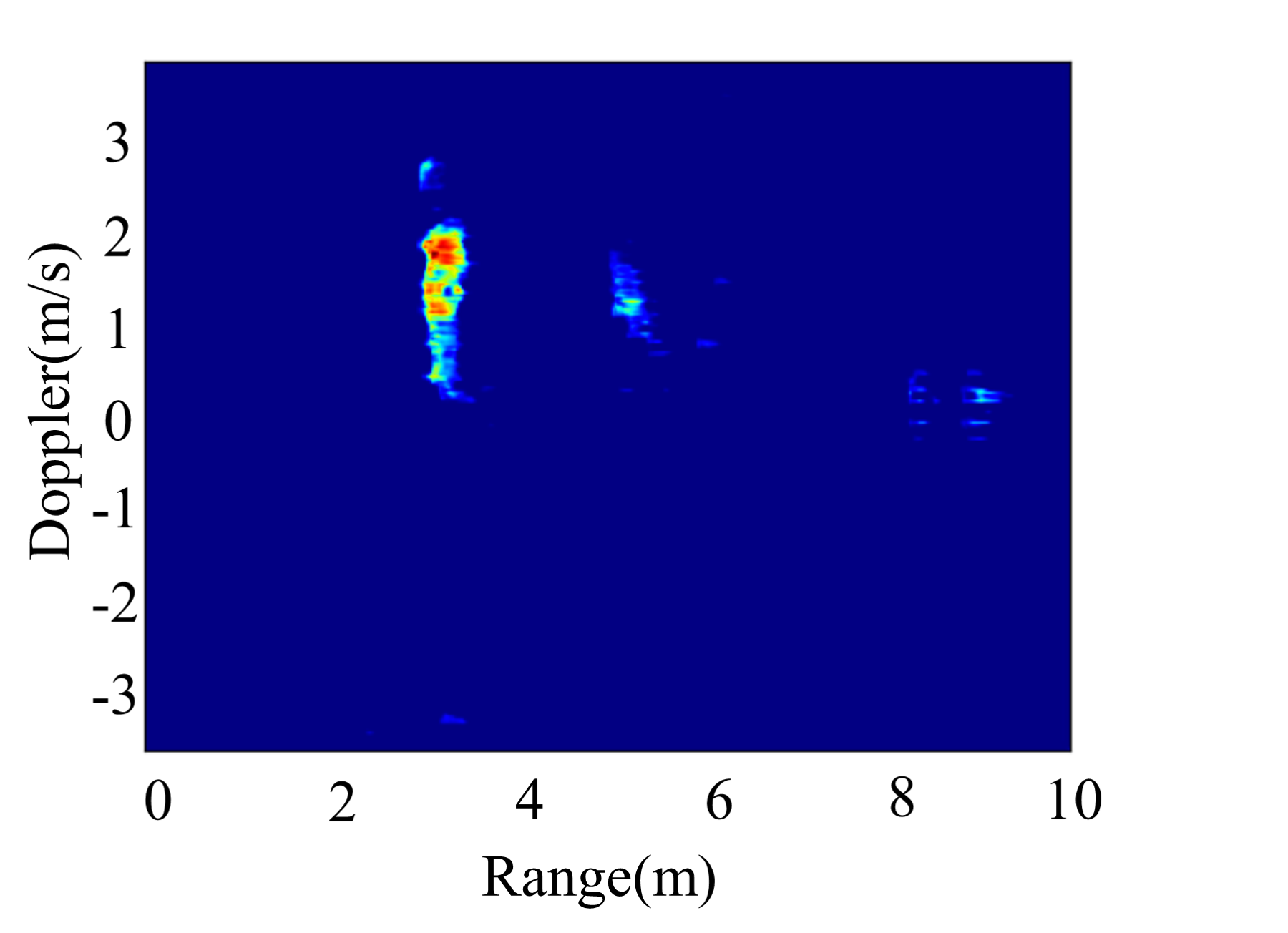}}
	\caption{Original RDM and generated RDM of pedestrian 4}
 \vspace{-0.4cm}
	\label{fig_6}
\end{figure}

\begin{figure}[!ht]
 \vspace{-0.4cm}
	\centering
	\subfloat[\scriptsize\normalfont Original TDS]{\includegraphics[width=0.24\textwidth]{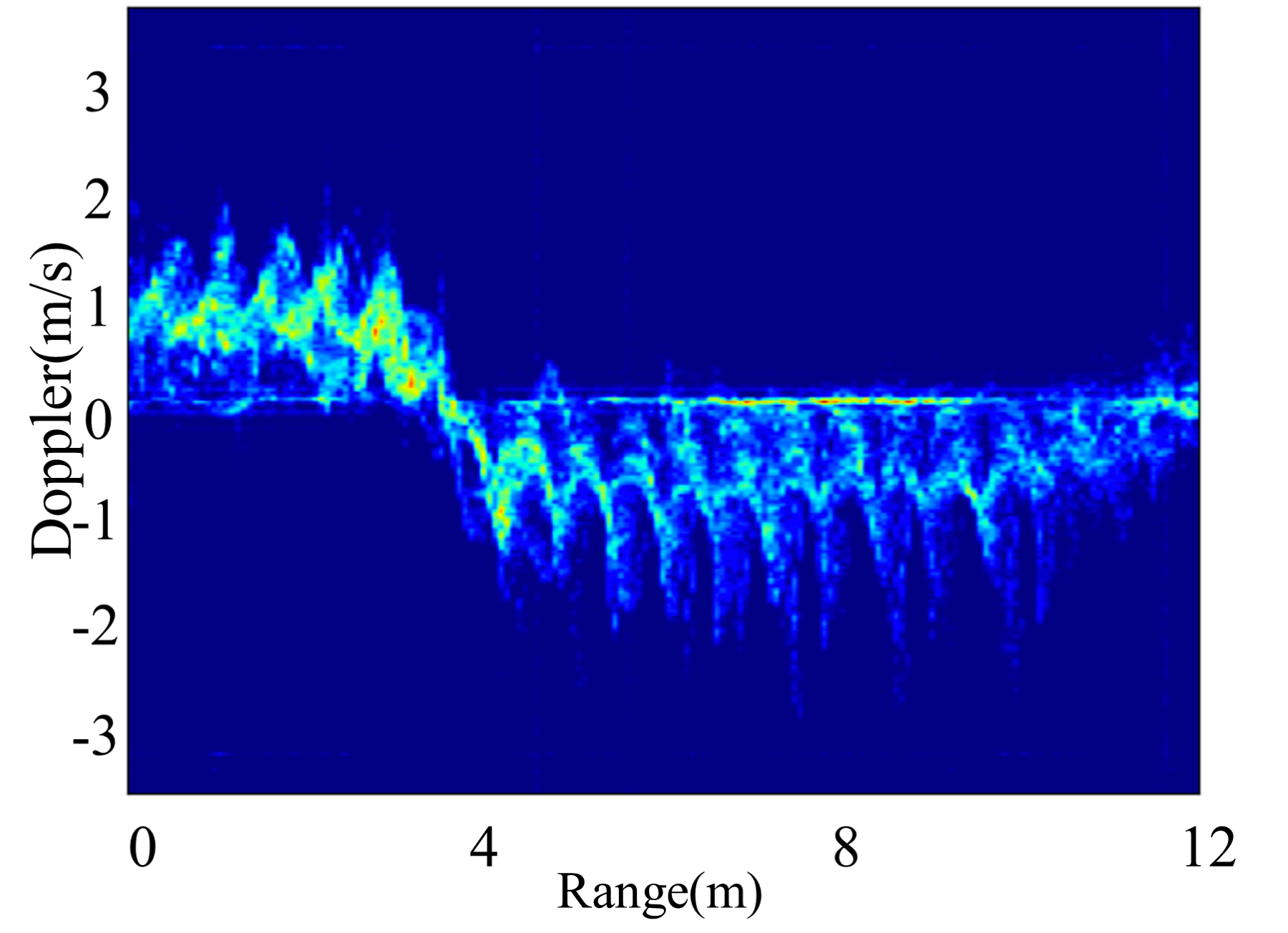}}
	\hfil
	\subfloat[\scriptsize\normalfont Generated TDS]{\includegraphics[width=0.24\textwidth]{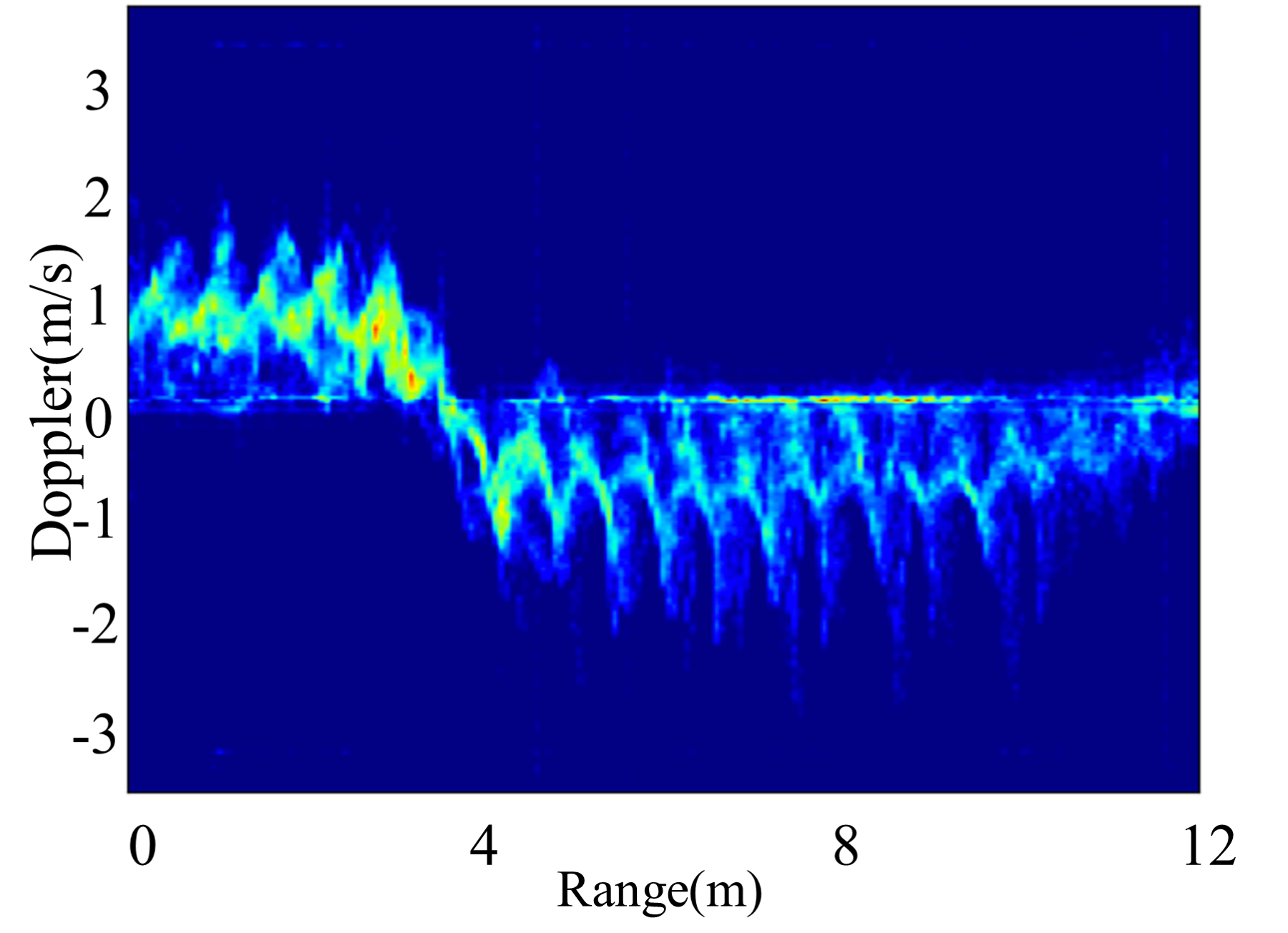}}
 	\caption{Original TDS and generated TDS of pedestrian 1}
	\label{fig_7}
\vspace{-0.3cm}
\end{figure}

Fig.\ref{fig_6} shows original RDM and generated RDM and Fig.\ref{fig_7} shows original TDS and generated TDS. Generated data has a high similarity with original data, and characteristics of original data are well learned.

\subsection{Results And Analysis of DEMCL Model}All results in this section are obtained by taking average value of five experimental results. 

\begin{figure}[!ht]
 \vspace{-0.5cm}
	\centering
	\subfloat[\scriptsize\normalfont Accuracy]{\includegraphics[width=0.24\textwidth]{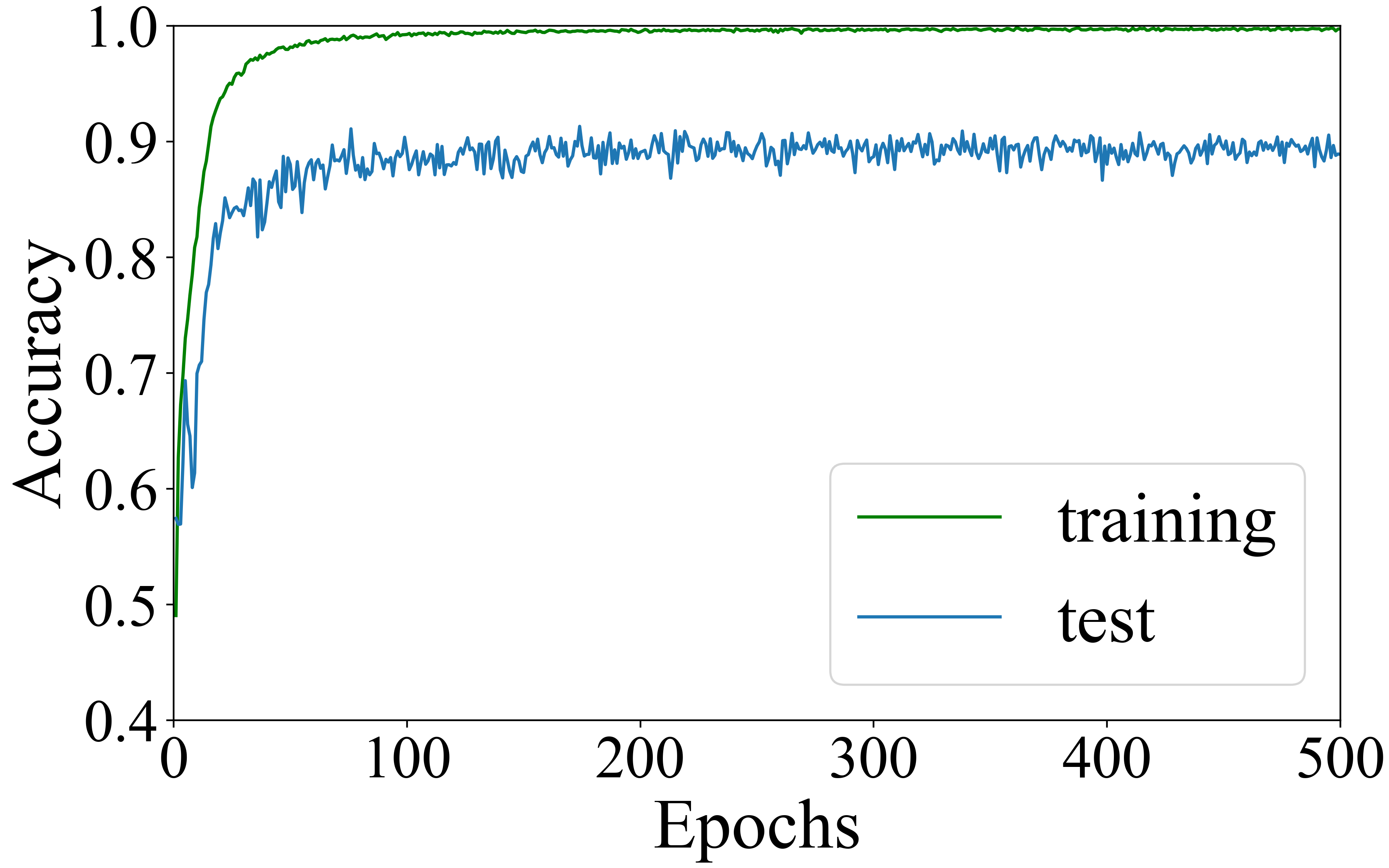}}
	\hfil
	\subfloat[\scriptsize\normalfont Loss]{\includegraphics[width=0.24\textwidth]{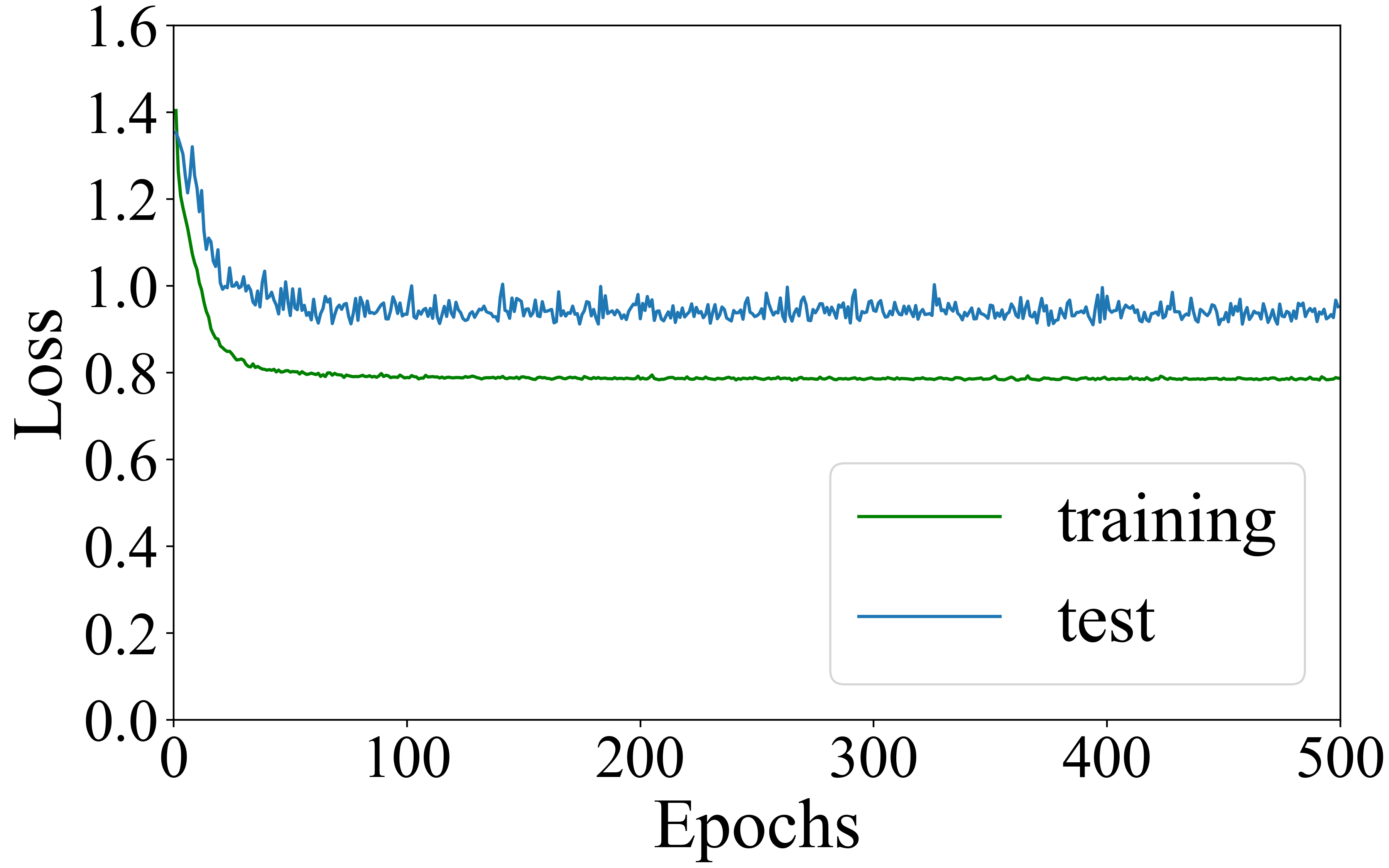}}
 	\caption{Accuracy and loss curves of training}
	\label{fig_8}
 \vspace{-0.3cm}
\end{figure}

DEMCL model obtains an accuracy rate of 92.83\%  for identified five pedestrians with test set (Fig.\ref{fig_8} (a)). The loss curves of training set and test set along with epochs tends to stabilize after $50$-th epoch (Fig.\ref{fig_8} (b)). 

\begin{figure}[!ht]
\vspace{-0.3cm}
	\centering
	\subfloat[\scriptsize\normalfont DEMCL]{\includegraphics[width=0.24\textwidth]{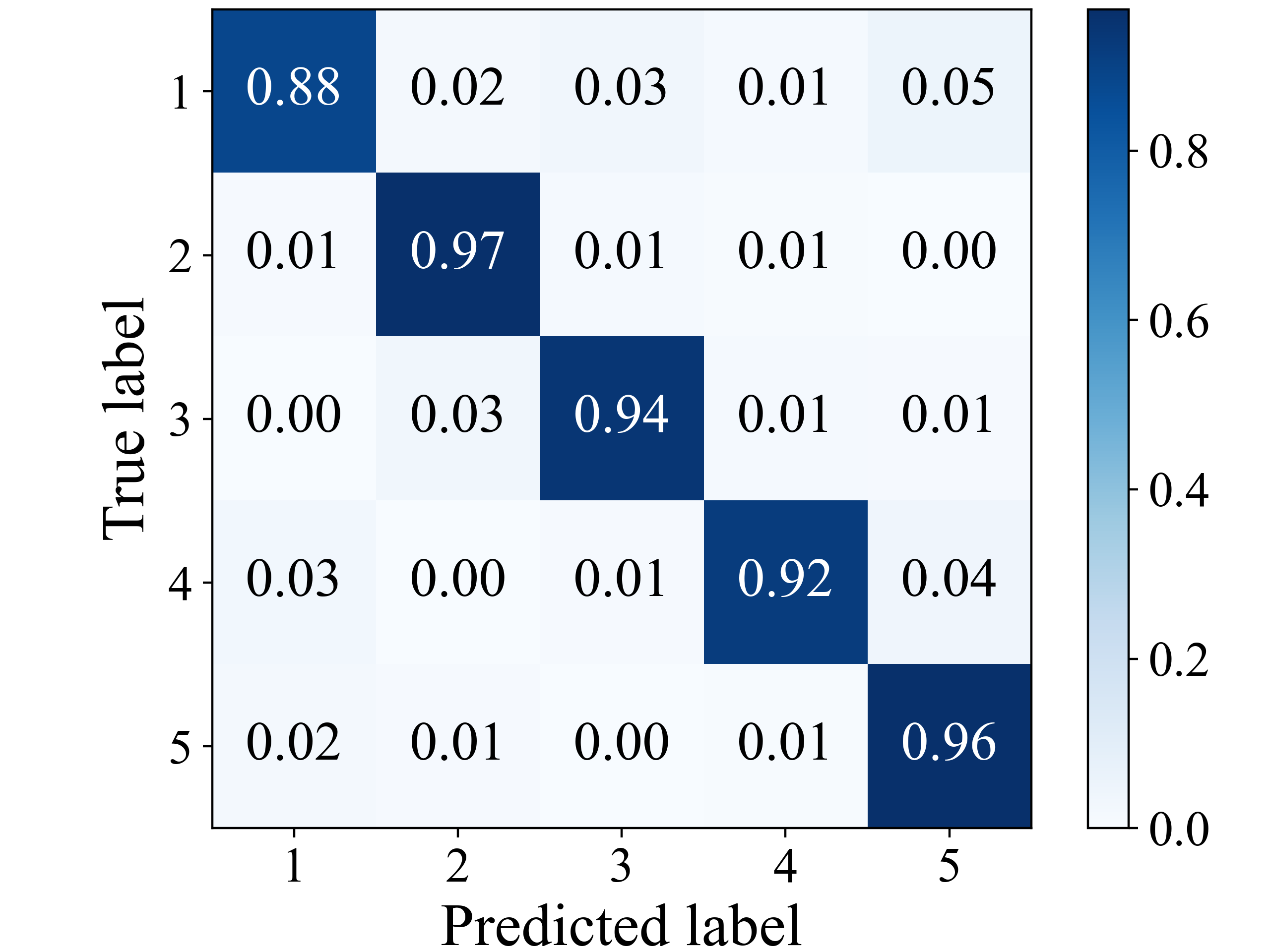}}
	\hfil
	\subfloat[\scriptsize\normalfont Xiang et al.]{\includegraphics[width=0.24\textwidth]{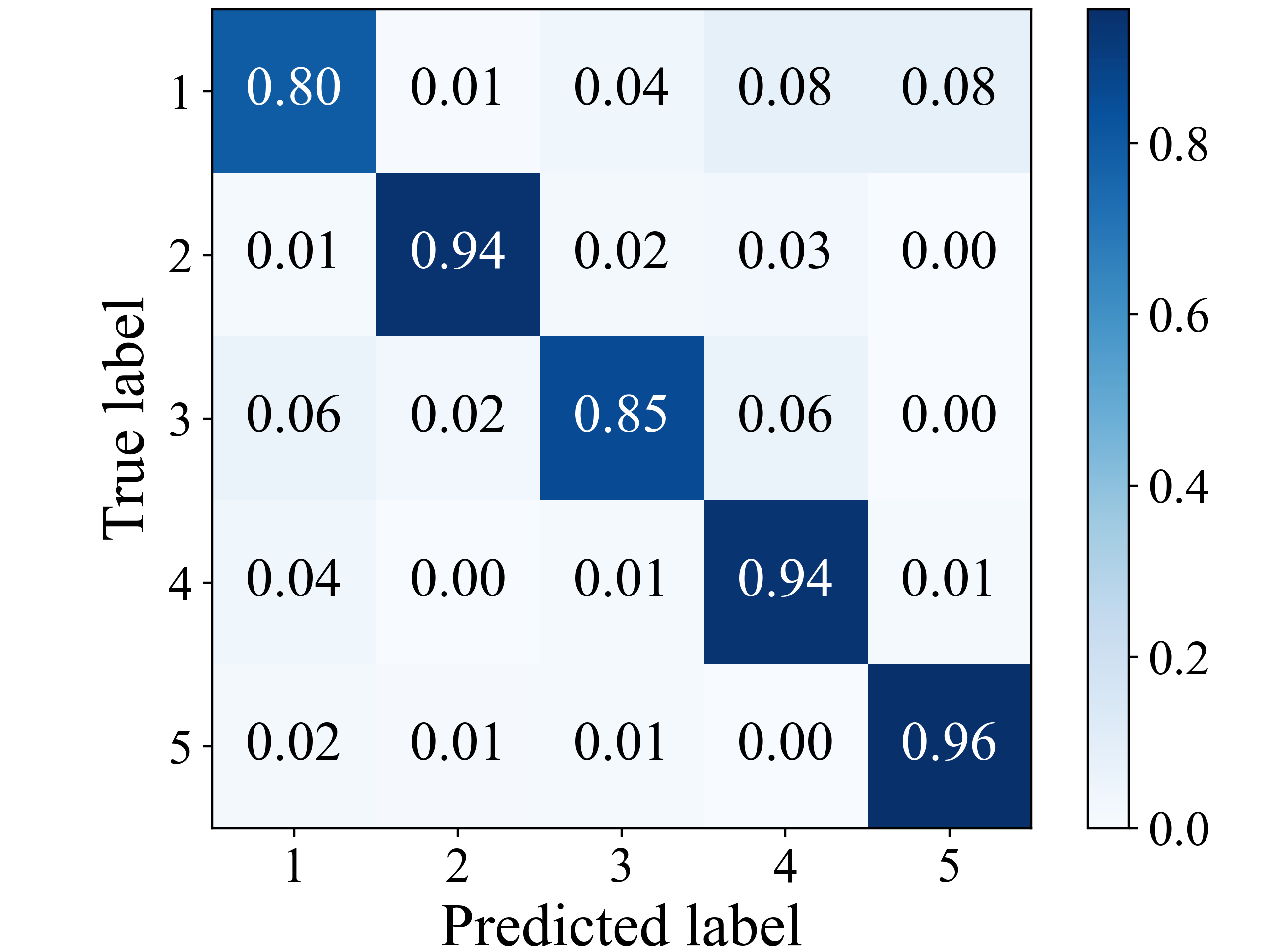}}
 	\caption{Confusion matrix comparison for five pedestrians in test set
  }
	\label{fig_9}
  \vspace{-0.3cm}
\end{figure}

Two confusion matrices (Fig.\ref{fig_9}) show the results from identification of five pedestrians with DEMCL model and method in \cite{xiang2022multi}, respectively.  DEMCL model has a high identification rate for all five pedestrians, even for pedestrian 1 with the lowest recognition accuracy rate, where the recognition accuracy rate exceeds 88\% (Fig.\ref{fig_9}. (a)). Compared to \cite{xiang2022multi}, the accuracy rate of pedestrian identification in pedestrians 1 and 3 has increased by about 8\% and 9\%, respectively. The rise is around 3\% for pedestrians 2. It proves that data enhancement is effective.

Next, DEMCL model is compared with other methodologies commonly used in the field of pedestrian identification. Comparison methods are summarized in Table \ref{table_3}. PyTorch toolkit is used to implement all methods. All methods are trained on the Tesla T4 for 500 epochs with a mini-batch size of 64.
\begin{table}[!th]
\setlength{\abovecaptionskip}{-0.1cm}
\setlength{\belowcaptionskip}{-0.3cm}
	\caption{ Comparison Methods \label{table_3}}
	\centering
	\resizebox{3in}{!}{
		\begin{tabular}{|c|c|}
			\hline
			Method & Network Description\\
			\hline
   			\multirow{2}{*}{Xiang et al. \cite{xiang2022multi} }& MCL: a novel multi-characteristic \\ & learning model with cluster\\
			\hline
			\multirow{2}{*}{Cao et al. \cite{cao2018radar}} & AlexNet, which has five convolution  \\ & layers  and two fully connected layers\\
			\hline
			Abdulatif et al. \cite{abdulatif2019person} & ResNet-50: a 50-layer deep residual network\\
			\hline
			\multirow{2}{*}{Lang et al. \cite{lang2020person}} & A plain convolutional neural network with \\ &
			a multi-scale feature aggregation strategy\\
			\hline   
   			\multirow{2}{*}{Vandersmissen et al. \cite{vandersmissen2018indoor} } &  A 6-layer CNN which has four convolutional \\ & layers and two  fully connected layers \\
			\hline
		\end{tabular}
	}
\end{table}
\vspace{-0.3cm}

\begin{figure}[!ht]
 \vspace{-0.3cm}
 \setlength{\abovecaptionskip}{0.cm}
	\centering
	\includegraphics[width=2.5in]{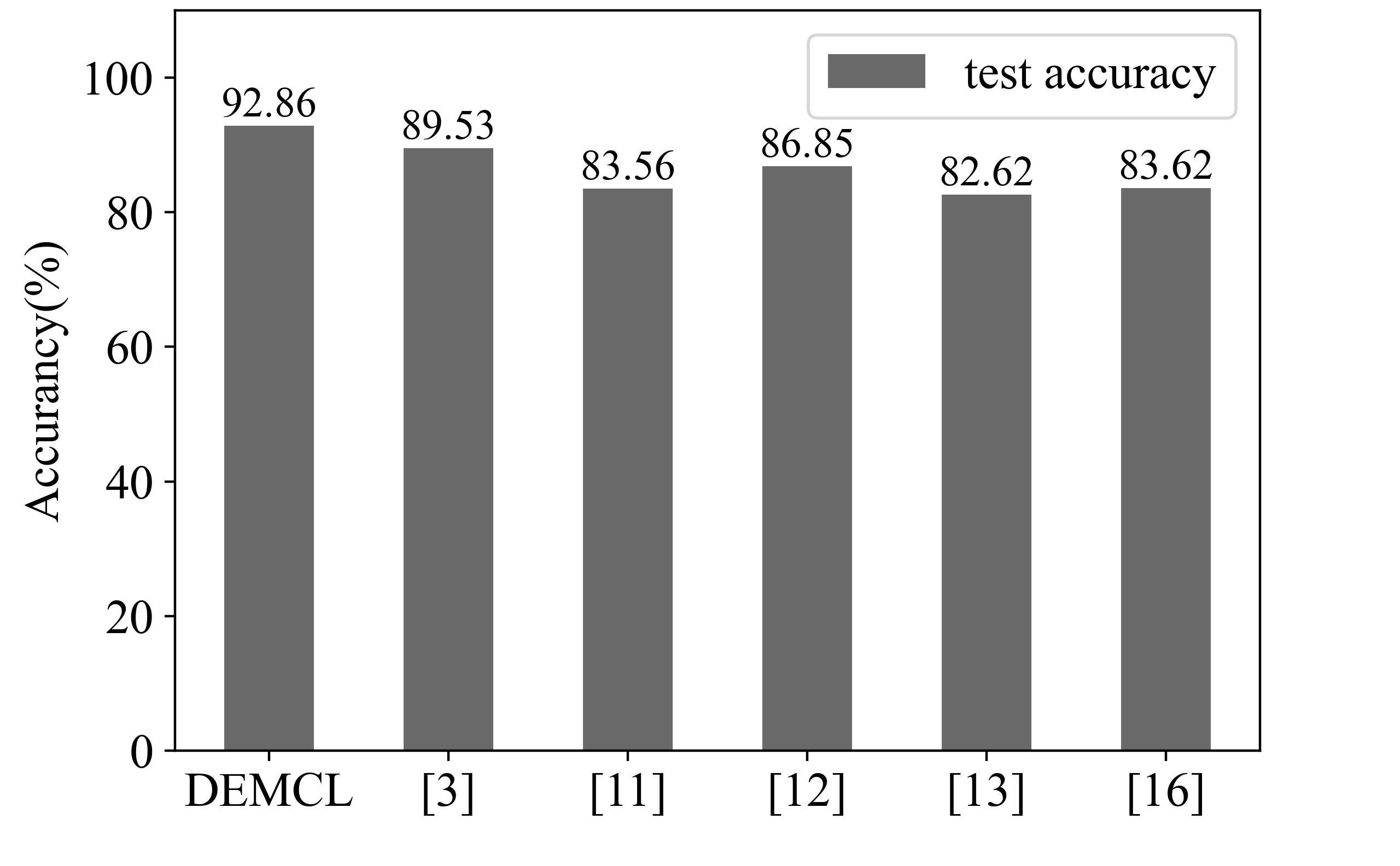}
	\caption{Accuracy rate comparison for pedestrian identification}
	\label{fig_10}
 \vspace{-0.3cm}
\end{figure}

As shown in Fig.\ref{fig_10}, The accuracy of DEMCL model is superior to various other methods in test set. The performance improved from 3.33\% to 10.24\% with test set.

\begin{figure}[!ht]
 \setlength{\abovecaptionskip}{0.cm}
	\centering
	\includegraphics[width=3.4in]{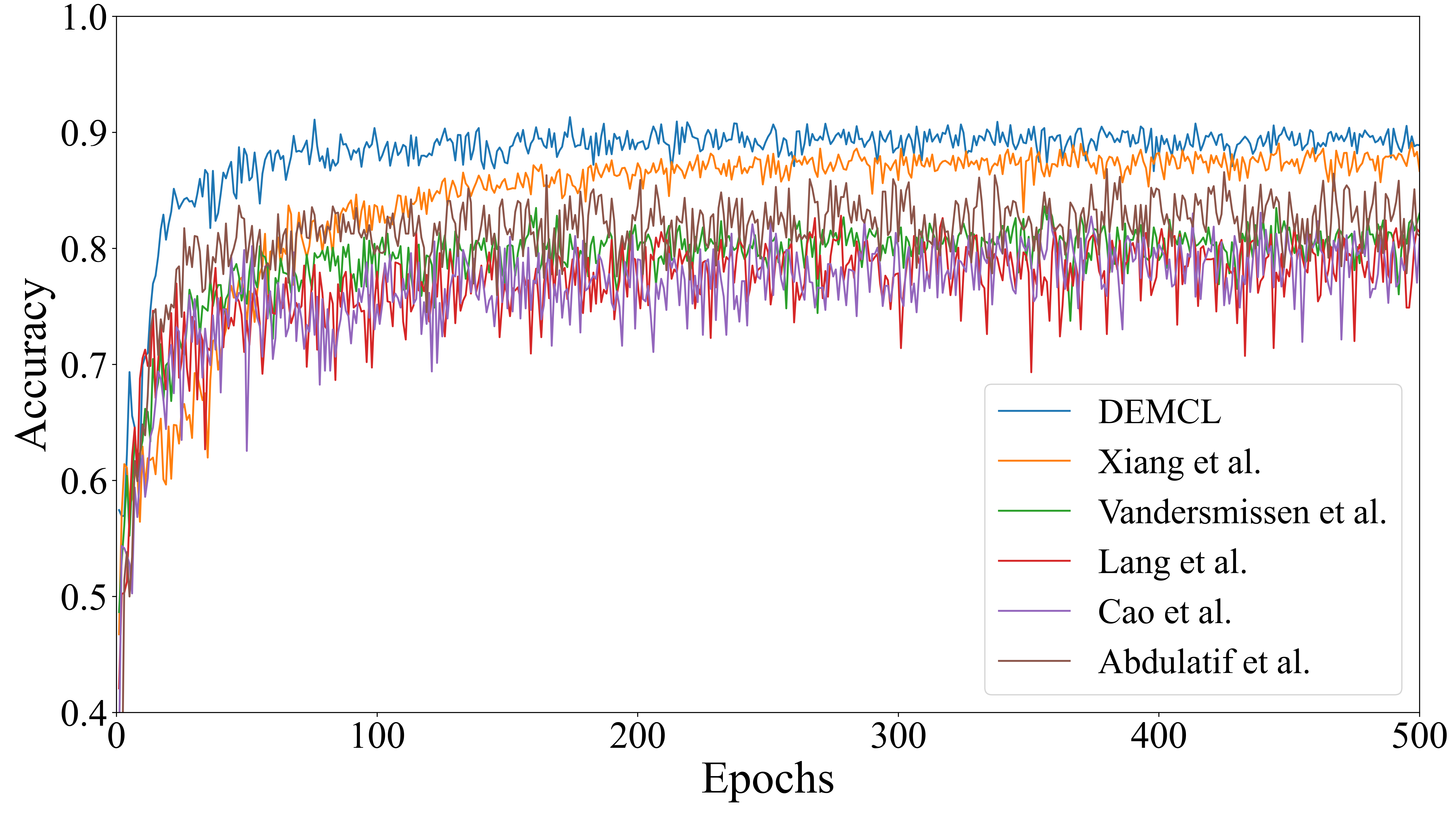}
	\caption{Pedestrian identification accuracy rate comparisons over epochs}
  \vspace{-0.3cm}
	\label{fig_11}
\end{figure}

Fig.\ref{fig_11} shows comparison of identification accuracy among all methods changed over epochs on the test set. DEMCL model converges fastest of all models and its accuracy is higher than other comparison methods along with epochs. In terms of stability, DEMCL model performs better than previous recognition algorithms.

\begin{figure}[!ht]

\setlength{\abovecaptionskip}{0.cm}
	\centering
	\includegraphics[width=2.5in]{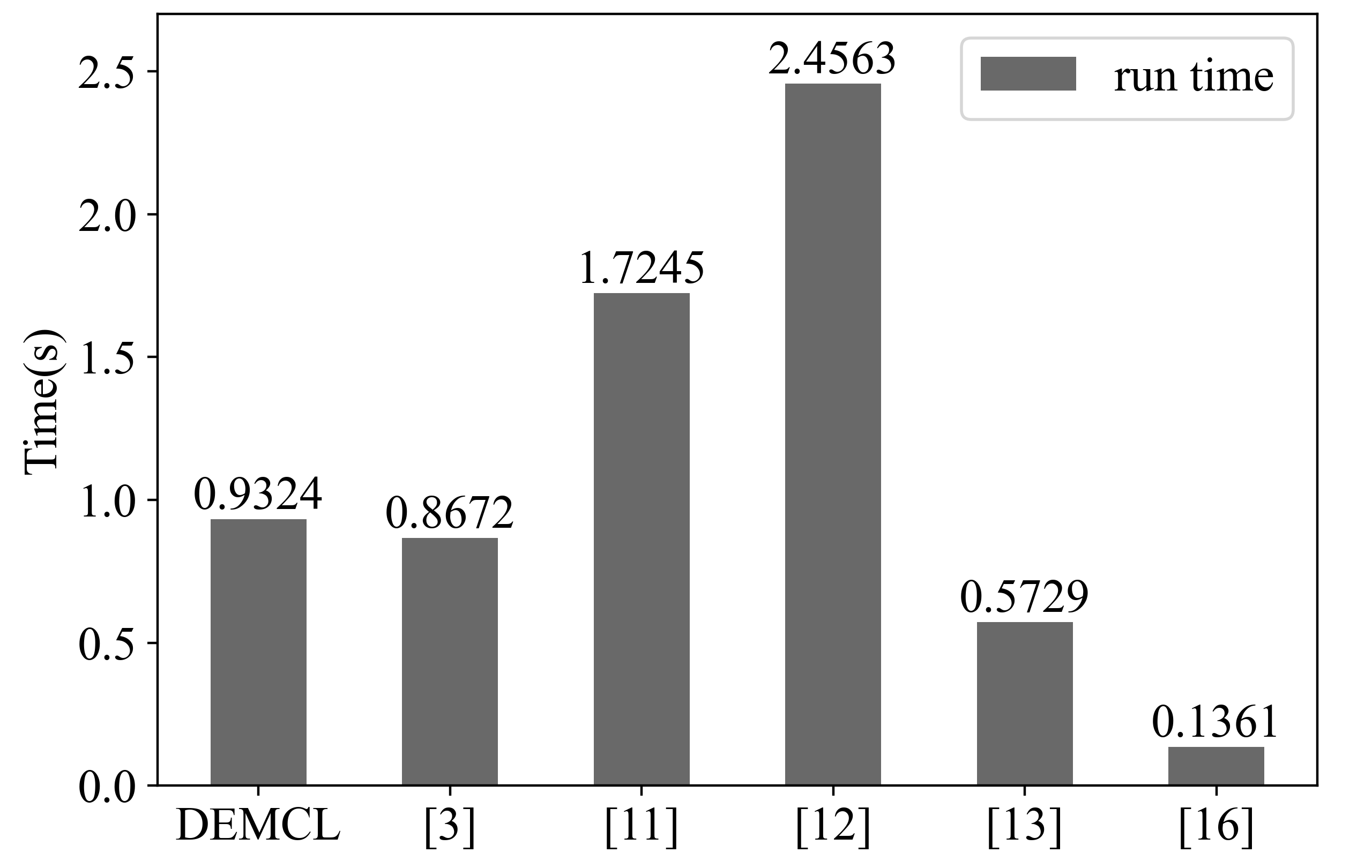}
	\caption{Pedestrian identification test time comparisons}
	\label{fig_12}
\end{figure}

Fig.\ref{fig_12} shows run time of all methods on the test set. DEMCL model has a short run time of 0.9324 seconds, which is less than the methods used in \cite{cao2018radar} and \cite{abdulatif2019person}, respectively. Run time of DEMCL model is longer than the methods in \cite{vandersmissen2018indoor} and \cite{lang2020person}, but its accuracy is 10.24\% and 9.24\% higher than their methods, respectively (Fig.\ref{fig_10}). Run time of method in \cite{xiang2022multi} is 0.0652 seconds less than DEMCL model's, but its accuracy is 3.33\% lower than DEMCL model's (Fig.\ref{fig_10}). Over all,  DEMCL model strikes an ideal balance between run time and accuracy.
\vspace{-0.4cm}
\section{Conclusions}

In this paper, we propose a novel data enhanced deep learning approach named DEMCL model to utilize more complementary information for pedestrian recognition. Experimental results show that our proposed model has an accuracy of 92.83\% on test set which is higher than other recognition methods. Moreover, DEMCL model performs more stable and provides better balance between time consumption and accuracy, make our model more practical. In the future, we will collect outdoor pedestrian datasets by a 77GHz FMCW radar platform on road environment. Then we may further improve this model by combining outdoor datasets with indoor datasets.

\end{document}